\documentclass[runningheads]{llncs}

\usepackage{floatrow}
\usepackage{algorithm}
\usepackage{algpseudocode}
\usepackage{amsmath,amsfonts,amssymb}
\usepackage[T1]{fontenc}
\usepackage{tikz}
\usetikzlibrary{automata, positioning, arrows}
\usepackage{graphicx}
\usepackage[english]{babel}

\begin{document}

\title{ALMA: Automata Learner using Modulo 2 Multiplicity Automata}

\author{Nevin George}

\authorrunning{N. George}

\institute{Yale University, New Haven CT 06511, USA\\
\email{nevin.george@yale.edu}}

\maketitle

\begin{abstract}
We present ALMA (Automata Learner using modulo 2 Multiplicity Automata), a Java-based tool that can learn any automaton accepting regular languages of finite or infinite words with an implementable membership query function. Users can either pass as input their own membership query function, or use the predefined membership query functions for modulo 2 multiplicity automata (M2MAs) and non-deterministic Büchi automata. While learning, ALMA can output the state of the observation table after every equivalence query, and upon termination, it can output the dimension, transition matrices, and final vector of the learned automaton. Users can test whether a word is accepted by performing a membership query on the learned automaton. \\

ALMA follows the polynomial-time learning algorithm of Beimel et al. (Learning functions represented as multiplicity automata. J. ACM 47(3), 2000). ALMA also implements a polynomial-time learning algorithm for strongly unambiguous Büchi automata by Angluin et al. (Strongly unambiguous Büchi automata are polynomially predictable with membership queries. CSL 2020), and a minimization algorithm for M2MAs by Sakarovitch (Elements of Automata Theory. 2009). \\

ALMA is unique from other similar tools in that hypotheses during the learning algorithm and the outputted learned automata are represented using M2MAs. The tool enables researchers to explore the practical and theoretical advantages of M2MAs, a relatively unexplored automaton representation which we argue is useful in the verification community for representing regular $\omega$-languages.

\end{abstract}

\section{Introduction} \label{Intro}

Angluin’s exact learning model 
\cite{Angluin87b} has been studied extensively in the context of learning theory, and it can be used to learn automata representing regular languages of finite and infinite words. In the model, a learner interacts with an oracle to learn a regular language using membership and equivalence queries. In a membership query, the learner learns from the oracle whether a word is in the language. In an equivalence query, the learner forms a hypothesis on what the language is, and the oracle either confirms that the hypothesis is correct or returns a counterexample, i.e, a word for which the hypothesis and language differ.

As one application of the exact learning model, Beimel et al. \cite{BeimelBBKV:2000} detail a polynomial-time algorithm to learn multiplicity automata using membership and equivalence queries. The algorithm can be generalized by replacing the membership query function for multiplicity automata with that of any other automaton accepting regular languages of finite or infinite words. We provide a high-level overview of the algorithm: the algorithm begins with a trivial hypothesis of the language, represented as a multiplicity automaton of dimension $1$ defined over some field $\cal K$. On each iteration of a loop, the algorithm performs an equivalence query. If the hypothesis is equivalent to the language, the algorithm terminates and outputs the hypothesis. Otherwise, the algorithm receives a counterexample from the oracle, which is used to improve the hypothesis. The algorithm terminates after $d$ iterations, where $d$ is the dimension of the smallest possible multiplicity automaton that can represent the language.

Strongly unambiguous Büchi automata (SUBAs) (defined in Section~\ref{automata}) are a type of non-deterministic Büchi automaton (NBA) first introduced by Bosquet and Löding \cite{BousquetL10}. SUBAs are useful for modeling reactive systems, as they are fully expressive, i.e., they can represent any regular $\omega$-language, and can often represent regular $\omega$-languages more succinctly than other NBA representations~\cite{AngluinAF20}. Angluin et al.~\cite{AngluinAF20} also showed that SUBAs are learnable in polynomial time, further increasing their importance.

In this paper, we present ALMA, a Java-based tool that implements the algorithm of Beimel et al. to learn any arbitrary automaton representing regular languages of finite or infinite words with an implementable membership query function. Hypotheses during the learning algorithm and outputted learned automata are represented using M2MAs. Membership query functions have been implemented for M2MAs and NBAs, and users can enter as input any arbitrary membership query function that 1) takes as input any possible word that can be formed from the given alphabet, and 2) outputs 0 or 1 to indicate whether the word is in the language. 

To improve the runtime of learning SUBAs, ALMA does not use the general learning algorithm of Beimel et al., but rather the SUBA learning algorithm of Angluin et al.~\cite{AngluinAF20}. Also when learning M2MAs, ALMA first implements the algorithm of Sakarovitch~\cite{SakarovitchBook2009} to minimize the M2MA before running the learning algorithm. This is because in the SUBA learning algorithm, converting the input SUBA into an equivalent M2MA incurs a quadratic increase size. Minimizing M2MAs before learning enables the algorithm to learn significantly larger SUBAs, especially because empirically the M2MAs have been seen to often minimize to M2MAs of much smaller dimension. The effect of the minimization algorithm on a sample of input SUBAs can be seen in Table~\ref{tab2} in Appendix~\ref{EE}.

\subsubsection{Usefulness/Novelty}
In the verification community, M2MAs can be useful for representing regular $\omega$-languages, as they are learnable in polynomial time and often relatively succinct. They are also useful for tasks such as model checking, since performing operations such as intersection, union, complementation, emptiness, and equivalence on M2MAs is cheap \cite{AngluinAFG22}. Through the minimization and learning algorithms, ALMA enables researchers in the verification community to easily test and verify these properties of M2MAs, promoting further exploration into these useful automata. As an example, ALMA was used in the paper by Angluin et al.~\cite{AngluinAFG22} to explore the suitableness of representing regular $\omega$-languages using M2MAs. The authors used ALMA to convert SUBAs, NBAs, and DBAs into equivalent M2MAs, and they compared the succinctness of these M2MAs with that of DFAs accepting the same language. ALMA can similarly be used by other researchers to gain insights into M2MAs and their benefits/drawbacks as compared to other representations.

ALMA is the first publicly available implementation of the novel SUBA learning algorithm by Angluin et al.~\cite{AngluinAF20}. Using ALMA, users can run the algorithm to learn any regular $\omega$-language. Since learned automata are represented as M2MAs, users can use the many desirable properties of M2MAs to gain insights into the initial SUBA and regular $\omega$-language. In addition, since membership query functions are often relatively easy to implement and ALMA already provides a membership query function for the general NBA case, the scope of what ALMA can learn is large, promoting its usefulness in a wide variety of settings.

Many tools such as ROLL ($\omega$-Regular Language Learning Library) \cite{YongXAYJ19} and libalf \cite{BolligKKLNP16} already exist that can learn automata representing regular languages of finite and infinite words. However, ALMA is the first tool that uses M2MAs to represent hypotheses and the learned automaton in the learning algorithm. ALMA's usefulness lies not with necessarily being the fastest tool available to learn regular languages, but with exploring the practical benefits of M2MAs and the features of the algorithms by Beimel et al., Angluin et al., and Sakarovitch.

\section{Useful Definitions}

\subsection{Finite and Büchi Automata} \label{automata}
Let $\Sigma$ be a finite alphabet. Then $\Sigma^*$ and $\Sigma^\omega$ are the sets of all finite and infinite words, respectively, that can be formed using elements from $\Sigma$. A finite language is a subset of $\Sigma^*$, and an $\omega$-language is a subset of $\Sigma^\omega$. If $w$ is a word in $\Sigma^*$ or $\Sigma^\omega$, let $|w|$ be the length of $w$ and $w[i]$ be the $i$'th character of $w$.

A finite-state automaton $A$ is represented as a tuple ($\Sigma, Q, I, \Delta, F$), where $\Sigma$ is the alphabet, $Q$ is the finite set of states, $I \subseteq Q$ is the set of initial states, $\Delta \subseteq Q \times \Sigma \times Q$ is the set of transitions, and $F \subseteq Q$ is the set of final states. The automaton $A$ is deterministic if every pair $(q, \sigma) \in Q \times \Sigma$ appears as the first two elements in at most one triple in $\Delta$.

A run on $A$ for a word $w$ is a series of states $q_0, q_1, \ldots \in Q$ such that $\forall i$ satisfying $1 \leq i \leq |w|$, $(q_{i-1}, w[i], q_i) \in \Delta$. A run for a finite word is final if it ends in a final state. For infinite words, a run is final if it passes infinitely often through a final state. A run is accepting if it is final and begins at an initial state. The automaton $A$ accepts the finite/infinite word $w$ if there exists an accepting run for $w$.

Non-deterministic finite automata (NFAs) and non-deterministic Büchi automata (NBAs) are automata accepting finite and infinite words, respectively. Deterministic finite automata (DFAs) are deterministic NFAs, and deterministic Büchi automata (DBAs) are deterministic NBAs. Unambiguous finite automata (UFAs) and unambiguous Büchi automata (UBAs) are NFAs and NBAs, respectively, for which every word has at most one accepting run. A UBA is a strongly unambiguous Büchi automaton (SUBA) if every word has at most one final run.

\subsection{Modulo 2 Multiplicity Automata} \label{M2MA}
Assume a field $\cal K$ and some dimension $d$. A multiplicity automaton $A$ is represented as a tuple ($\Sigma, v_I , \{\mu_\sigma\}_{\sigma \in \Sigma}, v_F$), where $\Sigma$ is the alphabet, $v_I$ is the initial vector, each $\mu_\sigma$ is a transition matrix, and $v_F$ is the final vector. $v_I$ and $v_F$ have dimension $d \times 1$, and each $\mu_\sigma$ has dimension $d \times d$. The set of states is all row vectors $v \in \{0, 1\}^d$, and the initial state is $v_I^\top$. For a given word $w = \sigma_1\sigma_2\ldots\sigma_n$, let $\mu(w) = \mu_{\sigma_1}\mu_{\sigma_2}\ldots\mu_{\sigma_n}$. The set of reachable states is all vectors of the form $v_I^\top\mu(w)$. Associated with the automaton $A$ is a function $f_A:\Sigma^* \to \mathcal{K}$, where $\forall w \in \Sigma^*$, \[f_A(w) = v_I^\top\mu(w)v_F.\] A modulo 2 multiplicity automaton (M2MA) is a multiplicity automaton where $\mathcal{K} =$ GF(2) and all calculations are done modulo 2. An M2MA $A$ accepts a word $w \in \Sigma^*$ if and only if $f_A(w) = 1$.

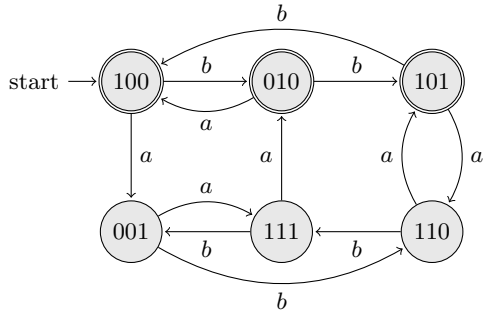
\begin{figure}
    \centering
    \begin{tikzpicture}[shorten >=1pt,node distance=2cm,on grid,auto]
  \tikzstyle{every state}=[fill={rgb:black,1;white,10}]

  \node[state,initial,accepting]  (s_0)                 {$100$};
  \node[state, accepting]                    (s_1) [right of=s_0]  {$010$};
  \node[state, accepting]                    (s_2) [right of=s_1]  {$101$};
  \node[state]                               (s_3) [below of=s_0]  {$001$};
  \node[state]                               (s_4) [right of=s_3]  {$111$};
  \node[state]                               (s_5) [right of=s_4]  {$110$};

  \path[->]
  (s_0) edge                node {$b$}  (s_1)
  (s_1) edge                node {$b$}  (s_2)
  (s_2) edge [bend right, above]  node {$b$}  (s_0)
  (s_1) edge [bend left]  node {$a$}  (s_0)
  (s_0) edge                node {$a$}  (s_3)
  (s_4) edge                node {$a$}  (s_1)
  (s_4) edge [below]        node {$b$}  (s_3)
  (s_5) edge [below]        node {$b$}  (s_4)
  (s_3) edge [bend left, above]  node {$a$}  (s_4)
  (s_5) edge [bend left]  node {$a$}  (s_2)
  (s_2) edge [bend left]  node {$a$}  (s_5)
  (s_3) edge [bend right, below]  node {$b$}  (s_5);
\end{tikzpicture}
    \caption{DFA for the M2MA $M$}
    \label{DFAforM}
\end{figure}

As an example, consider the following M2MA $M$ adapted from Angluin et al.~\cite{AngluinAF20}.\[M = (\{a, b\}, \begin{pmatrix} 1 & 0 & 0\end{pmatrix}^\top, \{\mu_a, \mu_b\}, \begin{pmatrix} 1 & 1 & 0\end{pmatrix}^\top\}\] where \[\mu_a = \begin{pmatrix} 0 & 0 & 1 \\ 1 & 0 & 0 \\ 1 & 1 & 1 \end{pmatrix} \text{and } \mu_b = \begin{pmatrix} 0 & 1 & 0 \\ 1 & 0 & 1 \\ 1 & 1 & 0 \end{pmatrix}.\] $M$ is equivalent to the DFA in Fig.~\ref{DFAforM}. We explain Fig.~\ref{DFAforM}: the computation begins at the initial state $\begin{pmatrix} 1 & 0 & 0\end{pmatrix}$. On reading an $a/b$ from the initial state, the DFA visits the states \[\begin{pmatrix} 1 & 0 & 0\end{pmatrix} \mu_a = \begin{pmatrix} 1 & 0 & 0\end{pmatrix}\begin{pmatrix} 0 & 0 & 1 \\ 1 & 0 & 0 \\ 1 & 1 & 1 \end{pmatrix} = \begin{pmatrix} 0 & 0 & 1\end{pmatrix}\]\[\begin{pmatrix} 1 & 0 & 0\end{pmatrix} \mu_b = \begin{pmatrix} 1 & 0 & 0\end{pmatrix}\begin{pmatrix} 0 & 1 & 0 \\ 1 & 0 & 1 \\ 1 & 1 & 0 \end{pmatrix} = \begin{pmatrix} 0 & 1 & 0\end{pmatrix}.\] Other states are visited similarly by multiplying the current state vector by $\mu_a$ or $\mu_b$. $\begin{pmatrix} 1 & 0 & 0\end{pmatrix}$, $\begin{pmatrix} 0 & 1 & 0\end{pmatrix}$, and $\begin{pmatrix} 1 & 0 & 1\end{pmatrix}$ are accepting states since \[\begin{pmatrix} 1 & 0 & 0\end{pmatrix}\begin{pmatrix} 1 & 1 & 0\end{pmatrix}^\top = \begin{pmatrix} 0 & 1 & 0\end{pmatrix}\begin{pmatrix} 1 & 1 & 0\end{pmatrix}^\top = \begin{pmatrix} 1 & 0 & 1\end{pmatrix}\begin{pmatrix} 1 & 1 & 0\end{pmatrix}^\top = 1,\] where $v_F = \begin{pmatrix} 1 & 1 & 0\end{pmatrix}^\top$.

M2MAs have many important properties described in Section~\ref{Intro} that make them useful in verification communities, e.g., learnable in polynomial time and cheap intersection, union, complementation, emptiness, and equivalence. For more information on M2MAs, we recommend reading the paper by Angluin et al.~\cite{AngluinAFG22}, which studies these properties of M2MAs extensively and performs a detailed analysis of M2MAs' ability to succinctly represent regular finite and $\omega$-languages.

\subsection{$L_\$$ Language} \label{LDollar}

Since Büchi automata accept infinite words and M2MAs accept finite words, Büchi automata and M2MAs cannot accept words from the same language $L$. However, Büchi \cite{Buchi62} showed that two regular $\omega$-languages are equivalent if and only if they agree on a set of ultimately periodic words, i.e., words of the form $u(v)^\omega$ where $u$ and $v$ are finite words and $v$ is non-empty. We then consider the language of finite words $L_\$ = \{u\$v \mid u(v)^\omega \in L\}$, which Calbrix et al. \cite{CalbrixNP93} showed is regular. If a Büchi automaton accepts an infinite language $L$, a finite automaton is said to also represent $L$ if it accepts $L_\$$. The $L_\$$ language is used by Angluin et al.~\cite{AngluinAF20} in their SUBA learning algorithm in order to obtain a finite automaton equivalent to the initial SUBA, and the learned M2MA accepts words from $L_\$$.

\section{Usage}

\subsection{Access and How to Run} 

ALMA is an open-source library freely available at the following GitHub repository: \url{https://github.com/nevingeorge/Learning-Automata}. The repository contains the executables, source files, and sample input files for ALMA, and the \textsf{README} document contains detailed information on how to use the executables and the required format for the input files. 

ALMA is a Java-based tool designed to be used on the command line. To run for example the executable \textsf{M2MA.jar}, a user should enter the command \textsf{java -jar M2MA.jar} within a Terminal/Command Prompt window. This will start the program, and the program will then print instructions on how to input the desired input file and flags.

\subsection{Executables}
ALMA consists of five main executables: 1) \textsf{M2MA.jar}, 2) \textsf{SUBA.jar}, 3) \textsf{minimize.jar}, 4) \textsf{NBA.jar},
and 5) \textsf{arbitrary.jar}. We provide a basic overview of each of the jar files and their use cases. For each executable, the output is always the dimension, final vector, and transition matrices of the learned/minimized M2MA. Also after the algorithms terminate, each executable allows users to check whether a word is accepted by the outputted M2MA. Infinite words are represented using the $L_\$$ language explained in Section~\ref{LDollar}.

\textsf{M2MA.jar} is used to minimize and learn M2MAs. The alphabet, dimension, final vector, and transition matrices of the input M2MA must be specified (the initial vector is always the vector with all zeros except for a 1 in the first row).

\textsf{SUBA.jar} is used to learn SUBAs using the algorithm of Angluin et al.~\cite{AngluinAF20}. The alphabet, number of states, final states, and transitions of the input SUBA must be specified (the only initial state is the first state).

\textsf{minimize.jar} is used to minimize M2MAs using the algorithm of Sakarovitch~\cite{SakarovitchBook2009}. The input is the same as that for \textsf{M2MA.jar}. \textsf{minimize.jar} can also be used to output the minimized M2MA equivalent to the input SUBA in the SUBA learning algorithm (i.e., it runs every step of the SUBA learning algorithm except for learning the final M2MA). In this case, the input is the same as that for \textsf{SUBA.jar}. 

\textsf{NBA.jar} is used to learn NBAs. The alphabet, number of states, final states, and transitions of the input NBA must be specified (the only initial state is the first state). \textsf{NBA.jar} uses approximate equivalence queries (described in Section~\ref{EQ}), which requires as input the number of tests to run, maximum length of a test, and maximum limit on the number of equivalence queries.

\textsf{arbitrary.jar} is used to learn arbitrary automata representing regular languages of finite and infinite words with an implementable membership query function. The membership query function must be defined in \textsf{MQ.java}. The only constraints on the function is that it must accept as input any possible word formed from letters in the alphabet, and it must output 0 or 1 depending on whether the word is contained in the language. \textsf{arbitrary.jar} also uses approximate equivalence queries, which requires the same input parameters as described for \textsf{NBA.jar}.

\subsection{Flags}
Users can enter the following optional flags to add or remove information from the output of the algorithm.

\begin{description}
   \item[-v:] The algorithm outputs the state of the observation table (a matrix consisting of the words whose membership in the language is known) after every equivalence query in the learning algorithm.
   
   \item[-m:] The algorithm outputs detailed information on the progress of the minimization algorithm. It gives status updates at different points in the algorithm, such as when it finishes creating the state/co-state spaces and the observation table. It also outputs the initial M2MA to be minimized and the final minimized observation table.
   
   \item[-d:] When running \textsf{minimize.jar} on a SUBA, the algorithm outputs only the dimension of the minimized M2MA instead of the dimension, final vector, and transition matrices of the M2MA.
   
   \item[-a:] After minimizing the M2MA, the algorithm outputs the number of states of the minimal deterministic finite automaton (DFA) that represents the same language as the M2MA.
\end{description}

\section{Implementation Details}

\subsection{Architecture} \label{arch}

\begin{figure}[ht]
    \centering
    \begin{tikzpicture}[
    node/.style={rectangle, draw=black, very thick, minimum size=15mm, text width=2cm, align=center}
    ]
    
    \node[node]      (1)                              {M2MA Input};
    \node[node]      (2)       [right=of 1] {Minimization Algorithm [Section \ref{min}]};
    \node[node]      (3)       [right=of 2] {Learning Algorithm [Section \ref{LA}]};
    \node[node]      (4)       [right=of 3] {Final Check [Section \ref{test}] and Output};
    
    \draw[->, very thick] (1.east) -- (2.west);
    \draw[->, very thick] (2.east) -- (3.west);
    \draw[->, very thick] (3.east) -- (4.west);
    \end{tikzpicture}
    \caption{\textsf{M2MA.jar} Architecture}
    \label{M2MA Structure}
\end{figure}

\begin{figure}[ht]
    \centering
    \begin{tikzpicture}[
    node/.style={rectangle, draw=black, very thick, minimum size=15mm, text width=1.9cm, align=center}
    ]
    
    \node[node]      (1)                              {SUBA Input};
    \node[node]      (2)       [right=of 1] {SUBA to UFA [Section \ref{SUBA_LA}]};
    \node[node]      (3)       [right=of 2] {UFA to M2MA [Section \ref{SUBA_LA}]};
    \node[node]      (4)       [right=of 3] {\textsf{M2MA.jar}};
    
    \draw[->, very thick] (1.east) -- (2.west);
    \draw[->, very thick] (2.east) -- (3.west);
    \draw[->, very thick] (3.east) -- (4.west);
    \end{tikzpicture}
    \caption{\textsf{SUBA.jar} Architecture}
    \label{SUBA Structure}
\end{figure}

\begin{figure}[ht]
    \centering
    \begin{tikzpicture}[
    node/.style={rectangle, draw=black, very thick, minimum size=15mm, text width=2.5cm, align=center}
    ]
    
    \node[node]      (1)                              {M2MA Input};
    \node[node]      (2)       [right=of 1] {Minimization Algorithm [Section \ref{min}]};
    \node[node]      (3)       [right=of 2] {Final Check [Section \ref{test}] and Output};
    
    \draw[->, very thick] (1.east) -- (2.west);
    \draw[->, very thick] (2.east) -- (3.west);
    \end{tikzpicture}
    \caption{\textsf{minimize.jar} Architecture}
    \label{Minimize Structure}
\end{figure}

\begin{figure}[ht]
    \centering
    \begin{tikzpicture}[
    node/.style={rectangle, draw=black, very thick, minimum size=15mm, text width=2.5cm, align=center}
    ]
    
    \node[node]      (1)                              {NBA/Arbitrary Input};
    \node[node]      (2)       [right=of 1] {Learning Algorithm [Section \ref{LA}]};
    \node[node]      (3)       [right=of 2] {Final Check [Section \ref{test}] and Output};
    
    \draw[->, very thick] (1.east) -- (2.west);
    \draw[->, very thick] (2.east) -- (3.west);
    \end{tikzpicture}
    \caption{\textsf{NBA.jar} and \textsf{arbitrary.jar} Architecture}
    \label{NBA Arbitrary Structure}
\end{figure}

Figures~\ref{M2MA Structure}-\ref{NBA Arbitrary Structure} provide a high-level overview of the architecture for each of the five executables. The learning algorithm of Beimel et al.~\cite{BeimelBBKV:2000}, described in more detail in Section~\ref{LA}, is implemented in every executable except for \textsf{minimize.jar}. The M2MA minimization algorithm of Sakarovitch~\cite{SakarovitchBook2009} (described in Section~\ref{min}) is implemented in \textsf{M2MA.jar}, \textsf{SUBA.jar}, and \textsf{minimize.jar}. All five executables first take in as input from the command line the name of the input file and optional flags. The input is passed to a parser, which then reads the input and sends the parameters of the automaton to be learned/minimized to the algorithms. The executables also all run final checks (described in Section~\ref{test}) on the outputted M2MA. Once the checks complete, users can test whether a word is accepted by the automaton.

\subsection{M2MA Minimization Algorithm} \label{min}

The M2MA minimization algorithm, implemented in the $\textsc{minimize}$ function of \textsf{M2MA.java}, is described in abstract terms in the work by Sakarovitch \cite{SakarovitchBook2009} and in more concrete detail in the appendix of \cite{AngluinAFG22}. The implemented minimization algorithm follows the pseudo-code in the appendix of \cite{AngluinAFG22} closely. 

The algorithm requires a heavy use of linear algebra, much of which is done using the Apache Commons Math package. While testing the executables, however, many of the unminimized M2MAs were seen to be relatively sparse with few 1's in the transition matrices. To take advantage of the sparseness, ALMA implements custom linear algebra functions that use a sparse representation of matrices. Matrix rows are represented using arrays containing the locations of the 1's in the row. For example, the row $\begin{bmatrix}0 & 1 & 0 & 0 & 0 & 1\end{bmatrix}$ is represented as $\begin{bmatrix}2 & 6\end{bmatrix}$, indicating that the row has a 1 in columns 2 and 6. Matrix multiplication, dot products, and tests for linear independence are implemented using the sparse matrix representation. Since the minimization function deals with M2MAs, the custom linear algebra functions perform all calculations modulo 2, which further improves the runtime.

\subsection{General Learning Algorithm} \label{LA}
The learning algorithm of Beimel et al.~\cite{BeimelBBKV:2000} is the core algorithm underpinning ALMA, and it is implemented in the \textsc{learn} function of \textsf{M2MA.java}. We provide a high-level overview of the algorithm in Section~\ref{Intro}, and more details on the algorithm are found in \cite{BeimelBBKV:2000}. Like the minimization algorithm, the linear algebra is implemented using a combination of the Apache Commons Math package and custom functions using the sparse representation for matrices. 

The default membership query function the learning algorithm uses is that for M2MAs. As described in Section~\ref{M2MA}, an M2MA $A = (\Sigma, v_I , \{\mu_\sigma\}_{\sigma \in \Sigma}, v_F)$ accepts a word $w$ if $v_I^\top\mu(w)v_{F} = 1$. Equivalence queries for M2MAs are easy, since as a by-product of running the minimization algorithm we get a complete observation table for the automaton. The equivalence query function checks whether the current hypothesis agrees with the M2MA being learned on every word in the observation table, as well as all possible one-letter extensions of the words. If they agree on every word and one-letter extension, the algorithm terminates; otherwise, the algorithm returns a word for which they disagree on as a counterexample.

\subsection{SUBA Learning Algorithm} \label{SUBA_LA}
The SUBA learning algorithm of Angluin et al.~\cite{AngluinAF20} works as follows: the input SUBA is first converted into an equivalent UFA using a simple construction by Bosquet and Löding~\cite{BousquetL10}. If the SUBA has $n$ states, the constructed UFA has size $2n^2 + n$. Next, the UFA is converted to an equivalent M2MA of the same size. Lastly, the M2MA is learned using the algorithm of Beimel et al.~\cite{BeimelBBKV:2000}. The algorithms for the SUBA to UFA and UFA to M2MA conversions are found in \textsf{SUBA.java}, and the outputted M2MA is sent to \textsf{M2MA.java} to be minimized and learned.

\subsection{Learning NBA and Arbitrary Automata}
In \textsf{M2MA.jar}, \textsf{minimize.jar}, and \textsf{SUBA.jar}, the learning algorithm runs membership and equivalence queries on M2MAs. The learning algorithm for \textsf{NBA.jar}, however, uses a membership query function designed specifically for NBAs which we detail in Appendix~\ref{NBA MQ}. \textsf{arbitrary.jar} passes to the learning algorithm a membership query function specified by the user from \textsf{MQ.java}. Also since the equivalence query function for M2MAs doesn't work in the general setting, \textsf{NBA.jar} and \textsf{arbitrary.jar} use approximate equivalence queries that rely on testing a random sample of words from the language. 

\subsubsection{Approximate Equivalence Queries} \label{EQ}

The approximate equivalence query function is defined in \textsf{arbitrary.java}. Along with the hypothesis, it requires two parameters $n$ and $l$ as input. The function generates $n$ random words of length at most $l$, and it tests whether the hypothesis and automaton being learned agrees on these words. Increasing $n$ improves the accuracy of the function at the expense of the runtime. In the input file, users also give a limit $m$ on the number of equivalence queries that can be run. Since there are no guarantees on the size of the learned M2MA, the parameter $m$ prevents the algorithm from running for an arbitrarily long length of time.

\subsection{Checks} \label{test}

The code performs checks at various points of the algorithm. If a check fails, the code throws an exception and terminates the program. Example checks include checking the validity of the input (e.g., correct number of transition matrices, matrices only contain 0's and 1's, etc.), whether a matrix is invertible before running a linear equation solver, and if the dimension of the minimized M2MA equals that of the final learned M2MA.

After learning an M2MA, the code checks whether the outputted M2MA agrees with the input automaton on the membership of a random sample of words. By default, the code generates 1000 random words of length at most 25, but these constants can be modified. To perform this check, the code uses the previously described membership query functions for M2MAs, NBAs, and arbitrary automata, as well as a membership query function for SUBAs described in a paper by Bosquet and Löding~\cite{BousquetL10}. Users can also run membership queries on the outputted M2MA to manually confirm whether the M2MA accepts a given word.

\section{Conclusion}

In Appendix~\ref{EE}, we perform an experimental evaluation of ALMA and analyze its practical capabilities. ALMA has limitations - for example, the runtime of \textsf{M2MA.jar} becomes impractical for random M2MAs of size much larger than 100, and ALMA doesn't implement an exact equivalence query function for learning NBAs and arbitrary automata. However, ALMA is highly efficient for most standard use cases, and it can be used to promote further research into M2MAs and their properties. For example, in the paper by Angluin et al.~\cite{AngluinAFG22}, ALMA is used to find the dimension of the minimum M2MA that can represent a regular $\omega$-language. Angluin et al. compare this dimension with the size of other finite automata that represent the same language to analyze the succinctness of the M2MA representation. Along with serving as a useful tool for investigating M2MAs, ALMA confirms the theoretical results of Beimel et al. \cite{BeimelBBKV:2000} and Angluin et. al \cite{AngluinAF20,AngluinAFG22}, and is the first publicly available tool that can be used to explore these learning algorithms.

\subsubsection{Acknowledgements} I would like to thank Dana Angluin, Timos Antonopoulos, and Dana Fisman for their help with this paper and all the feedback and advice they gave. Dana Angluin especially helped me significantly during the entire process of creating ALMA, and I would like to thank her greatly for her support and mentorship.

\bibliographystyle{splncs04}
\bibliography{main}

\newpage
\appendix
\section{Appendix}

\subsubsection{NBA Membership Query Function} \label{NBA MQ}

The NBA membership query function is described in Algorithm~\ref{NBA MQ Alg}. As explained in Section~\ref{automata}, an NBA accepts a word if there exists a path for the word that begins at an initial state and passes infinitely often through a final state. The input parameters $u,v$ represent a word $w = u\$v$ in the $L_\$$ language.

\begin{algorithm}
\caption{NBA Membership Query Function}\label{NBA MQ Alg}
\begin{algorithmic}[1]
\Function{main}{$u$, $v$}
	\State $S_u \gets$ states reachable from the initial state on reading $u$ \label{start1}
	\State $S_{uv} \gets S_u \cup$ \Call{reachable}{$S_u$, $v$} \label{end1}
    \ForAll{$s \in S_{uv}$} \label{start2}	 \State $S_{uv}' \gets$ \Call{reachable}{$\{s\}$, $v$} \label{reach_loop1}
        \If{$s \in S_{uv}'$ $\And$ \text{passed a final state to reach} $s$}
            \State \Return 1
        \EndIf
    \EndFor
    \State \Return 0 \label{end2}
\EndFunction
\\
\Function{reachable}{$S$, $v$}
    \State $S_v \gets$ states reachable from a state in $S$ on reading $v$
    \State $S_v' \gets \varnothing$
    \Repeat \label{repeat}
        \State $S_v \gets S_v \cup S_v'$
        \State $S_v' \gets$ states reachable from a state in $S_v$ on reading $v$ \label{reach_loop}
    \Until{$S_v' \subseteq S_v$}
    \State \Return $S_v$
\EndFunction
\end{algorithmic}
\end{algorithm}

In Algorithm~\ref{NBA MQ Alg}, the \Call{reachable}{} function finds all the states reachable from a set of initial states $S$ on reading some positive number of $v$'s. On every iteration of the loop defined in line \ref{repeat}, the function finds the states reachable on reading another $v$, and the function terminates after an iteration of the loop where no new states are found. 

In the \Call{main}{} function, in lines \ref{start1}-\ref{end1} Algorithm \ref{NBA MQ Alg} stores in $S_{uv}$ all the states reachable from the initial state on reading a $u$ and a non-negative number of $v$'s. Then in lines \ref{start2}-\ref{end2}, the algorithm determines whether there is an accepting loop on any of the states in $S_{uv}$ (i.e., a path from a state in $S_{uv}$ to itself that passes through a final state).

\subsubsection{Runtime Analysis} Let $n$ be the number of states, $m$ be the number of transitions, and $l$ be the length of $u\$v$. Line \ref{start1} runs in $O(nml)$ - we read each of the $O(l)$ characters in $u$ one at a time, and with each character we calculate the new states that can be reached using the $O(m)$ transitions from the $O(n)$ states reached so far. \Call{reachable}{} runs in $O(n^2ml)$ - the loop runs at most $n$ times since there are at most $n$ states to add to $S$, and line \ref{reach_loop} runs in $O(nml)$. The loop in line \ref{start2} terminates after $O(n)$ iterations, and since \Call{reachable}{} is called in line \ref{reach_loop1}, the loop runs in $O(n^3ml)$. Therefore, Algorithm~\ref{NBA MQ Alg} runs in $O(n^3ml)$.

\section{Experimental Evaluation} \label{EE}

The GitHub repository contains many input files that can be used to test each of the executables. There exist input files for many different sizes of M2MAs, SUBAs, and NBAs, as well as input files for different edge cases (e.g., M2MAs of dimension 1).

\begin{table}[ht]
\begin{center}
\caption{\textsf{M2MA.jar} Runtime}
\label{tab1}
\begin{tabular}{| c | c |}
\hline
M2MA Dimension & Average Runtime\\
\hline
10& 0.33s\\
\hline
20& 2.44s\\
\hline
30& 9.76s\\
\hline
40& 28.57s\\
\hline
50& 82.57s\\
\hline
60 & 177.00s\\
\hline
70 & 355.09s\\
\hline
80 & 657.82s\\
\hline
90 & 1084.85s\\
\hline
100 & 1819.80s\\
\hline
\end{tabular}
\end{center}
\end{table}

\begin{table}[ht]
\begin{center}
\caption{\textsf{SUBA.jar} Runtime}
\label{tab2}
\begin{tabular}{| c | c | c | c | c |}
\hline
SUBA Language & SUBA Size & \begin{tabular}{@{}c@{}}Unminimized \\ M2MA Dim\end{tabular} & \begin{tabular}{@{}c@{}}Learned \\ M2MA Dim\end{tabular} & Runtime\\
\hline
$a \Sigma^* (\Sigma^* b \Sigma^*)^\omega$ & 2 & 10 & 5 & 0.20s\\
\hline
$\Sigma^* a \Sigma^5 a b^\omega$ & 8 & 136 & 10 & 0.30s\\
\hline
\begin{tabular}{@{}c@{}}$((a+b)^* (a (a+b) a (a+b) c$ \\ $+ b (a+b) b (a+b) d))^\omega$\end{tabular} & 9 & 171 & 92 & 50.86s\\
\hline
$(a^* a^4 b)^\omega$ & 5 & 55 & 32 & 0.57s\\
\hline
$(a^* a^5 b)^\omega$ & 6 & 78 & 44 & 2.04s\\
\hline
$(a^* a^6 b)^\omega$ & 7 & 105 & 58 & 6.72s\\
\hline
$a^\omega$ & 1 & 3 & 3 & 0.05s\\
\hline
$(ab^5)^\omega$ & 6 & 78 & 43 & 0.45s\\
\hline
$(ab^{10})^\omega$ & 11 & 253 & 133 & 25.11s\\
\hline
$(ab^{15})^\omega$ & 16 & 528 & 273 & 411.48s\\
\hline
$(ab^{20})^\omega$ & 21 & 903 & 463 & 3064.32s\\
\hline
\end{tabular}
\end{center}
\end{table}

\begin{table}[ht]
\begin{center}
\caption{\textsf{NBA.jar} Runtime}
\floatfoot{*The learned M2MA dimensions and runtimes for \textsf{NBA.jar} can vary widely due to the inherent randomness.}
\label{tab3}
\begin{tabular}{| c | c | c | c |}
\hline
NBA Size & \begin{tabular}{@{}c@{}}Number of \\ tests/EQ\end{tabular} & \begin{tabular}{@{}c@{}}Learned M2MA \\ Dimension\end{tabular} & Runtime\\
\hline
2 & 10000 & 3 & 1.02s\\
\hline
4 & 10000 & 7 & 0.25s\\
\hline
6 & 10000 & 24 & 1.80s\\
\hline
8 & 100000 & 27 & 14.00s\\
\hline
10 & 100000 & 83 & 983.39s\\
\hline
\end{tabular}
\end{center}
\end{table}

Tables~\ref{tab1}-\ref{tab3} give a sense of the runtimes for \textsf{M2MA.jar}, \textsf{SUBA.jar}, and \textsf{NBA.jar}. The experiments were run on a standard laptop with 8 GB RAM, 8 cores, and a CPU frequency of 3200 MHz. 

Table~\ref{tab1} details the runtime of \textsf{M2MA.jar}. Fifty random M2MAs with the alphabet $\{a, b, c\}$ were generated for each of the dimensions $10, 20, \ldots, 50$, and the average runtimes were calculated. For the dimensions $60, 70, \ldots, 100$, ten random M2MAs were generated. These results can be reproduced using the executable \textsf{M2MA\_experiments.jar} in the Experimental Evaluation folder of the GitHub repository. Instructions on how to use the jar file are printed to standard output once the executable is run.

Table \ref{tab2} details the runtime of \textsf{SUBA.jar}. The SUBA input files used to generate the results are found in the Experimental Evaluation folder of the GitHub repository.

Table \ref{tab3} details the runtime of \textsf{NBA.jar}. The NBAs in the table were randomly generated, with one NBA generated per size. Since \textsf{NBA.jar} uses approximate equivalence queries, the table contains a column for the number of tests to run for every equivalence query. The maximum length of a test for every NBA in the table is 25. The NBA input files used to generate the results are found in the Experimental Evaluation folder of the GitHub repository.

\subsubsection{Practical Capabilities}

The runtimes for \textsf{M2MA.jar} increase at a roughly cubic rate. For smaller M2MAs, the program runs quickly, but for M2MAs of dimension larger than 100, the program can take hours to terminate. For the purpose of using ALMA to explore the properties of M2MAs, it is unlikely that one will need to work with M2MAs of dimension much larger than 100, so the program's runtime should not pose a significant constraint. 

One may think that experimenting with the SUBA learning algorithm, which incurs a $2n^2+n$ blow up in size going from the initial SUBA to the converted M2MA, will be impractical when dealing with SUBAs of size even greater than 10. However, as can be seen in Table~\ref{tab2}, the unminimized M2MA in the SUBA learning algorithm often minimizes to a much smaller M2MA, which is why the tool implements the minimization algorithm of Sakarovitch~\cite{SakarovitchBook2009}. For example, the SUBA of size 21 representing the language $(ab^{20})^\omega$ converts into a large unminimized M2MA of dimension 903. However, it minimizes to an M2MA of dimension 463, and \textsf{SUBA.jar} runs in less than an hour for this SUBA.

The learned M2MA dimensions and runtimes for \textsf{NBA.jar} depend heavily on the number of tests performed every equivalence query. To get decent approximations on the learned M2MA dimension for NBAs of size larger than 10, the number of tests per equivalence query should be at least on the order of 10 to the 5th or 6th power. One can try to change the maximum length of the tests to get better results. For randomly generated NBAs of size much larger than 10, the trade-off between the runtime and accuracy becomes much more apparent, and the number of tests per equivalence query may have to be decreased for the runtime to be practical.

\end{document}